\documentclass[prl,aps,floatfix,showpacs,tightenlines,twocolumn,superscriptaddress,amsmath,amssymb]{revtex4}

\usepackage{graphicx}% Include figure files
\usepackage{dcolumn}% Align table columns on decimal point
\usepackage{bm}% bold math
\usepackage{epsfig}

\begin{document}

\title{Hybrid Quantum Computation in Quantum Optics}

\author{P.\ van Loock}\email{vanloock@nii.ac.jp}
\affiliation{National Institute of Informatics, 2-1-2
Hitotsubashi, Chiyoda-ku, Tokyo 101-8430, Japan}

\author{W.\ J.\ Munro}
\affiliation{National Institute of Informatics, 2-1-2
Hitotsubashi, Chiyoda-ku, Tokyo 101-8430, Japan}
\affiliation{Hewlett-Packard Laboratories, Filton Road, Stoke
Gifford, Bristol BS34 8QZ, United Kingdom}

\author{Kae Nemoto}
\affiliation{National Institute of Informatics, 2-1-2
Hitotsubashi, Chiyoda-ku, Tokyo 101-8430, Japan}

\author{T.\ P.\ Spiller}
\affiliation{Hewlett-Packard Laboratories, Filton Road, Stoke
Gifford, Bristol BS34 8QZ, United Kingdom}

\author{T.\ D.\ Ladd}
\affiliation{National Institute of Informatics, 2-1-2
Hitotsubashi, Chiyoda-ku, Tokyo 101-8430, Japan}
\affiliation{Edward L. Ginzton Laboratory, Stanford University,
Stanford, California 94305-4088, USA}
% \affiliation{Nanoelectronics
% Collaborative Research Center, IIS, University of Tokyo, Tokyo 153-8505, Japan}

\author{Samuel L.\ Braunstein}
\affiliation{Computer Science, University of York, York YO10 5DD,
United Kingdom}

\author{G.\ J.\ Milburn}
\affiliation{Centre for Quantum Computer Technology, Department of
Physics, University of Queensland, Australia}

\begin{abstract}
We propose a hybrid quantum computing scheme where qubit degrees
of freedom for computation are combined with quantum continuous
variables for communication. In particular, universal two-qubit
gates can be implemented deterministically through qubit-qubit
communication, mediated by a continuous-variable bus mode
(``qubus''), without direct interaction between the qubits and
without any measurement of the qubus. The key ingredients are
controlled rotations of the qubus and unconditional qubus
displacements. The controlled rotations are realizable through
typical atom-light interactions in quantum optics. For such
interactions, our scheme is universal and works in any regime,
including the limits of weak and strong nonlinearities.
\end{abstract}


\maketitle

%\section{Introduction}

There are various proposals for realizing quantum computers
\cite{fort00,spiller05}. At the few-qubit level, some proposals
have been demonstrated already in the laboratory. These
proof-of-principle demonstrations include schemes based on, for
instance, trapped ions \cite{cirzol95}, linear optics
\cite{klm01,kok06} and nuclear spins in liquid-state molecules
\cite{nmr97}. For the long-term prospects of scalability,
``solid-state'' qubits are also of great interest. For their
realization, the toolbox and all the fabrication and manufacturing
expertise developed for conventional IT could be exploited.
However, at present, such solid-state-based schemes lag behind the
other approaches and are at best at the one- or two-qubit
demonstration level.

For processing photonic qubits directly in an optical quantum
computer, the large Kerr-type nonlinearities needed for a
two-qubit gate are hard to obtain with single photons. A possible
way to circumvent this obstacle is to apply only linear
transformations, supplemented by measurement-induced
nonlinearities \cite{klm01,kok06}. The simplest forms of these
linear-optical gates have been realized already \cite{kok06}.
There are also proposals that combine the advantages of the
solid-state and the optical approaches; the main idea of these
schemes is to use single photons as a bus to mediate interactions
between non-nearest neighbours of solid-state qubits
\cite{cir97,bos99,dua03,bro03,lim05,barrett05}. In principle, this
enables one to add arbitrarily many qubits to a system, in order
to achieve universality and scalability. Two-qubit gates can be
achieved for any pair and there is no need for the qubits to be so
close together such that individual addressing is no longer
possible.
%\cite{footnote1}.

Significant difficulties with single-photon-based buses arise due
to the demanding requirements on the generation and detection of
the photons. In particular, successful near-deterministic gate
performance depends on efficient detectors that unambiguously
detect a single photon. As a result, with typically low practical
detector efficiencies  the gates will be highly nondeterministic.
However, efficient local gates are essential ingredients in, for
example, long-distance quantum communication via quantum repeaters
\cite{Briegel98}. In such schemes, inefficient gates require more
expensive quantum resources. In addition, measurement-based gates
are typically slow, limited by the measurement speed. It is
therefore desirable to circumvent the need for measurements.

All of the above-mentioned proposals for realizing a quantum
computer rely exclusively on discrete variables (DV). The quantum
information is encoded into qubits (actual, or effective---a 2D
subspace in a larger Hilbert space) and, in some cases, qubits are
also used as a bus to mediate interactions. This includes the
original ion-trap proposal \cite{cirzol95} where the two lowest
states of a vibrational mode mediate a gate between two ion-qubits
(based on two internal ion states). There are now also efficient
and practical approaches to quantum communication based on {\it
continuous variables} (CV) \cite{bra05}. Inspired by these
results, and in order to avoid both direct qubit-qubit
interactions and the use of single photons, here we propose the
following ``hybrid quantum computer'': universal two-qubit gates
shall be achieved indirectly through the interaction between the
qubits and the quadrature phase amplitude of a common bosonic
mode. The CV mode plays the role of a communication bus which we
call a ``qubus''. This approach brings together the best of both
worlds, utilizing DV for processing and CV for communication.

The idea of the CV qubus computer has been applied to ion traps
\cite{mil99,mol99,sor00,mil00} and other systems \cite{Spiller},
but here we focus on a {\it quantum optical} realization. In this
approach, the qubits are either atomic or photonic, and the qubus
is an electromagnetic field mode; the CV are the phase-space
variables of this field mode. Although efficient homodyne
detection of certain phase-space variables (quadratures) is
possible, no measurement will be needed in our scheme. By design,
under ideal conditions, the bus mode disentangles automatically
from the qubits after a sequence of interactions.
Measurement-induced errors are thus avoided and the gates become
deterministic, requiring neither measurement-result-dependent
post-selection nor any feed-forward operations on the qubits.
Moreover, we make no assumptions about the strength of the
qubit-qubus interactions; our scheme works in any regime,
including the limits of weak and strong nonlinearities. The
proposal here relies on two new important concepts: the {\it
exact} simulation of controlled phase-space displacements via
controlled rotations and uncontrolled displacements; and an {\it
efficient} all-cavity implementation of this simulation.

In contrast to existing CV-mediated proposals for measurement-free
ion-trap gates based upon conditional displacements
\cite{mil99,mol99,sor00,mil00}, our proposed two-qubit gate
is based on {\it conditional rotations}.
These are obtainable from the fundamental Jaynes-Cummings interaction
$\hbar g (\sigma^{-}a^{\dagger} + \sigma^{+}a)$ in the
dispersive limit \cite{Schleich}, which gives
\begin{equation}
\label{Hint} H_{\rm int} = \hbar \chi \sigma_z a^{\dagger} a\,.
\end{equation}
Here, $a$ ($a^{\dagger}$) refers to the annihilation (creation)
operator of an electromagnetic field mode in a cavity and
$\sigma_z = |0\rangle\langle 0| - |1\rangle\langle 1|$ is the
corresponding qubit operator from the set of Pauli operators
$\{\sigma_x,\sigma_y,\sigma_z\}$ for a two-level atom in the
cavity (with ground state $|0\rangle$ and excited state
$|1\rangle$) \cite{footnoteatom}. The atom-light coupling strength
is determined via the parameter $\chi=g^2/\Delta$, where $2g$ is
the vacuum Rabi splitting for the dipole transition and $\Delta$
is the detuning between the dipole transition and the cavity
field. The Hamiltonian in Eq.~(\ref{Hint}) generates a conditional
phase-rotation of the field mode, dependent upon the state of the
atomic qubit. Note that the dispersive interaction for a
high-fidelity conditional rotation does not require strong
coupling; the only requirement is a sufficiently large
cooperativity parameter \cite{Ladd2006}.

It has been pointed out \cite{wan01} that a suitable set of
Hamiltonian terms, including conditional rotations and
unconditional displacements $\{\sigma_x a^{\dagger} a,\sigma_z
a^{\dagger} a,x\}$, is, in principle, sufficient for universal
quantum computation. Here our main concern is how to efficiently
utilize these universal resources. Throughout, we use the
definition for quadrature operators $X(\phi)=(a^{\dagger} e^{i
\phi} + a e^{-i \phi})$ such that $X(0)\equiv x$ and
$X(\pi/2)\equiv p$ play the roles of ``position'' and
``momentum'', respectively, with $[x,p]= 2i$ for
$[a,a^{\dagger}]=1$. We now demonstrate how a universal two-qubit
gate can be
 implemented via the Jaynes-Cummings-type interaction from
Eq.~(\ref{Hint}) and additional unconditional displacements.

\begin{figure}[!htb]
\includegraphics[width=5.5cm]{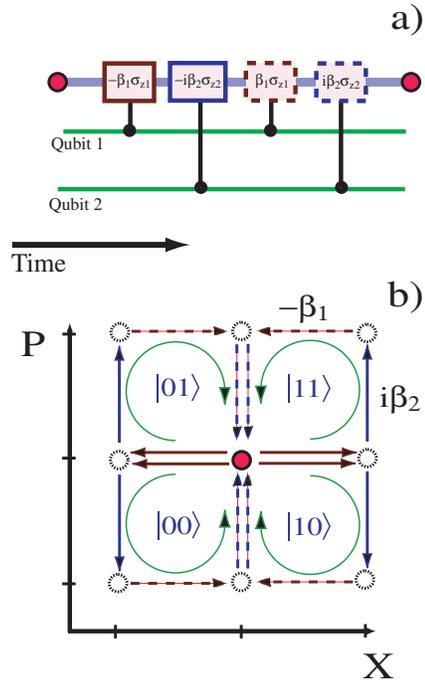}
\caption{\label{fig1} a) Circuit diagram of a universal two-qubit
gate based on controlled displacements between the qubits and the
probe bus. b) Schematic phase space evolution
of a coherent qubus amplitude (with the $\beta$'s chosen real),
depending on the four basis states of
the two qubits.}
\end{figure}

Our two-qubit gate relies upon the basic principle that a CV
mode acquires a phase shift whenever it goes along a closed loop
in phase space. This phase shift only depends on the area of the
loop and not on its form \cite{wan01} and it originates
from the fact that for any sequence of two displacements, the
total displacement operator contains an extra phase factor,
\begin{eqnarray}\label{twodisplacements}
D(\beta_1) D(\beta_2)= \exp \left[ i\,{\rm Im}\left(\beta_1
\beta_{2}^*\right) \right] D(\beta_1 +\beta_2) \; .
\end{eqnarray}
Here $D(\beta) = \exp(\beta a^{\dagger} - \beta^* a)$ is the usual
quantum optical displacement operator. In this sense such a
two-qubit gate can be regarded as a geometric phase gate
\cite{wan01}. In Ref.~\cite{Spiller}, it was shown how a
conditional phase gate on qubits can be realized by creating
almost closed loops in phase space through controlled rotations
and uncontrolled displacements. However, this gate is imperfect,
as even under ideal conditions the CV qubus does not disentangle
completely from the qubits, leading to an intrinsic dephasing
error. Here, instead of directly applying the interaction in
Eq.~(\ref{Hint}) to create a closed path, we instead {\it simulate
controlled displacements via the controlled rotations} in
Eq.~(\ref{Hint}). With controlled displacements available it is
straightforward to implement a conditional phase gate, as we now
describe.

Let us assume that an arbitrary two-qubit state enters the gate
such that the total initial state (of the two-qubit-qubus system)
may be written as
\begin{equation}
\label{Psiinitial}
\left(c_1|00\rangle + c_2 |01\rangle +
c_3|10\rangle + c_4 |11\rangle\right)|{\rm qubus}\rangle\,,
\end{equation}
with a qubus-probe mode initially in an arbitrary state $|{\rm
qubus}\rangle$. The two-qubit gate follows from four conditional
displacements. The sequence of operations is shown in
Fig.~\ref{fig1}a). This defines the total unitary operator
\begin{eqnarray}\label{totalunitary}
U_{\rm tot} \equiv D(i\beta_2 \sigma_{z2}) D(\beta_1 \sigma_{z1})
D(-i\beta_2 \sigma_{z2}) D(-\beta_1 \sigma_{z1}).
\end{eqnarray}
Using Eq.~(\ref{twodisplacements}), it is straightforward to show
that
\begin{eqnarray}\label{totalunitaryresult}
U_{\rm tot} = \exp \left[2 i\; {\rm Re}(\beta_1^* \beta_2)\;\sigma_{z1}
\sigma_{z2}\right] \;.
\end{eqnarray}
Apparently, when this operator acts on the two-qubit-qubus system,
the only effect is the generation of phase factors conditional on
the two-qubit state. Although it is entangled with the qubits
during the gate, the qubus mode finishes in its initial state,
disentangled from the qubits. The evolution does not depend on
this qubus state---a convenient choice would be a coherent state
\cite{footnote2}.

For the case of real $\beta_1$ and $\beta_2$, the effect of the
total operation on a bus coherent state, conditional on the state
of the qubits, is illustrated in Fig.~\ref{fig1}b). By choosing
$\beta_1 \beta_2 = \pi/8$, a total initial state as in
Eq.~(\ref{Psiinitial}) gives a final pure two-qubit state of
\begin{eqnarray}\label{cphase-gate-exact}
e^{-i\frac{\pi}{4}} \;U\otimes U \left(c_1|00\rangle + c_2
|01\rangle + c_3|10\rangle - c_4 |11\rangle\right)\,,
\end{eqnarray}
where $U \equiv e^{i \frac{\pi}{4} \sigma_z}$. Thus, up to a
global phase and local unitaries, we obtain a controlled-phase
gate.

So far we have assumed that we can perform conditional
displacements in order to construct the operation $U_{\rm tot}$
of Eq.~(\ref{totalunitary}). In quantum optics, it is hard to
generate such conditional displacements directly through
photon-atom or photon-photon interactions. However, the
Jaynes-Cummings-type interaction of Eq.~(\ref{Hint}) is readily
available. We now show that this interaction is
sufficient to generate the required conditional displacements. More
specifically, we will use a series of pulses and interactions of
the type of Eq.~(\ref{Hint}) in order to effectively simulate a
controlled displacement. No approximations will be needed for this
purpose, so our method is applicable to any regime
of the interaction in Eq.~(\ref{Hint}), including the weak and the
strong coupling limits.

\begin{figure}[!t]
\includegraphics[width=5.5cm]{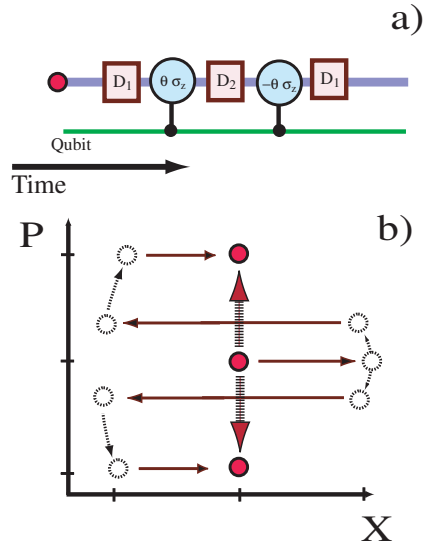}
\caption{\label{fig2} a) Circuit diagram for an effective
controlled displacement constructed from uncontrolled
displacements and controlled rotations. b) Schematic phase space
evolution of a coherent qubus amplitude during the controlled
displacement.}
\end{figure}

We define conditional rotations as generated by Eq.~(\ref{Hint}),
with an effective interaction time $\chi t\equiv \theta$.
Consider the following operation,
\begin{eqnarray}\label{exactseries}
\mathcal U \equiv D(\alpha\cos\theta)e^{-i \theta \sigma_z
a^{\dagger} a}D(-2\alpha)e^{i \theta \sigma_z a^{\dagger}
a}D(\alpha\cos\theta),
\end{eqnarray}
consisting of unconditional displacements and conditional
rotations. Using $e^{-i \theta a^{\dagger} a} a \,e^{i \theta
a^{\dagger} a} = a e^{i \theta}$, hence $e^{-i \theta a^{\dagger}
a} D(\alpha) \,e^{i \theta a^{\dagger} a} = D(\alpha e^{-i
\theta})$, and the rule in Eq.~(\ref{twodisplacements}), we find
that the sequence in Eq.~(\ref{exactseries}) {\it exactly}
realizes a conditional displacement such that
\begin{eqnarray}\label{exactseriesresult}
\mathcal U = D\left(2i \alpha \,\sin\theta\,\sigma_z\right)\;.
\end{eqnarray}
Figure~\ref{fig2} illustrates the sequence of uncontrolled
displacements and controlled rotations to simulate a controlled
displacement.

The resultant operation in Eq.~(\ref{exactseriesresult})
corresponds to a conditional displacement by $2i
\alpha\sin\theta$. The entire sequence of Eq.~(\ref{totalunitary})
can now be achieved through uncontrolled displacements and
controlled rotations of the probe via the Jaynes-Cummings-type
interaction from Eq.~(\ref{Hint}). This provides an exact
mechanism to create the controlled phase gate. Assuming $\beta_1 =
\beta_2 = \sqrt{\pi/8}$, the strength of the conditional rotations
for simulating the conditional displacements are determined by the
parameter $d\equiv 2 |\alpha| \sin\theta=\sqrt{\pi/8}\approx 0.6$.
For example, with a Jaynes-Cummings coupling and interaction time
corresponding to $\theta\sim 10^{-2}$, unconditional displacements
of about $|\alpha|^2\sim 10^4$ photons are needed. However, we may
also satisfy $d\approx 0.6$ using strong nonlinearities,
$\theta\sim\pi/2$ with weak qubus displacements of the order
$|\alpha|\sim 1$.

We shall now compare the controlled phase gate proposed here to
the one described in Ref.~\cite{Spiller}. Two crucial differences
exist, both of which highlight the advantages of the new gate.
\begin{itemize}
\item First and foremost the gate of Ref.~\cite{Spiller}
is only approximate. It has an intrinsic error since the qubus
probe does not completely disentangle from the qubits, causing a
dephasing effect on the qubits. To keep this error small requires
$|\alpha|\theta^2 \ll 1$, so the gate only works when $\theta \ll
1$. The gate presented here does not have this limitation. In this
sense, our scheme here is universal and can be applied to various
physical systems, in any coupling regime.
\item The second difference is important from a practical point of
view and relates to the local single qubit rotations needed to
realize the gate in (\ref{cphase-gate-exact}). The gate in
Ref.~\cite{Spiller} requires single qubit rotations of the form
$e^{i |\alpha|^2 \theta \sigma_z}$. This places considerable
sensitivity on $\alpha$ and $\theta$, requiring them to be known
accurately enough to perform single qubit operations that scale as
$|\alpha|^2 \theta$. In the gate presented here we only require a
unitary of the form $e^{i \frac{\pi}{4} \sigma_z}$, which is
independent of both $\alpha$ and $\theta$ and thus much less
demanding.
\end{itemize}

In order to accomplish the sequence $U_{\rm tot}$ in
Eq.~(\ref{totalunitary}) via the operation $\mathcal U$ from
Eq.~(\ref{exactseries}), it appears to be necessary to couple the
qubus mode out of the cavity and back into it whenever an
unconditional displacement must be applied via an external local
oscillator field. However, this rather inefficient feature can be
avoided in an {\it all-cavity-based} implementation of $\mathcal
U$. A very natural way to generate the unconditional displacements
is to drive the qubus mode directly with a classical pump. Such
driving can be represented by the Hamiltonian $H_d= \hbar \epsilon
X(\phi)$, with $\epsilon$ real, effectively resembling a
phase-space displacement. For instance, with $\phi=0$, our system
Hamiltonian is of the form $H(\epsilon,\chi \sigma_z)= \hbar
\epsilon \left( a^\dagger+ a\right) + \hbar \chi a^\dagger a
\sigma_z$. Now applying this operation $U(\epsilon,\chi
\sigma_z)=\exp[-\frac{i}{\hbar} H(\epsilon,\chi \sigma_z) t]$ for
a time $t$ followed by $U(\epsilon,-\chi \sigma_z)$
\cite{signchange} for the same time $t$ implements an effective
controlled displacement of the form $D_{\sigma_z}=D[\frac{2
\epsilon}{\chi}(1-e^{i \chi t \sigma_z})]$ which for $\chi t \ll
1$ has the more usual form $D[ 2 i \epsilon t \sigma_z]$. This
conditional operation $D_{\sigma_z}$ also contains unconditional
operations, but this does not effect the operation of the gate
\cite{purecontrolleddispalcement}. In fact, these unconditonal
displacements are undone by further conditional operations and so
our controlled displacement can be reduced to just two operations.
The entire two-qubit gate then requires only eight operations.

A further issue we need to consider is the robustness of our
two-qubit gate against noise and errors, for example caused by
photon losses in the qubus mode \cite{jeong06}. This is
particularly important, as we suggest to place the two atomic
qubits in two different cavities in order to avoid the
complication of individually addressing more atoms in one cavity.
A simple loss model reflects part of the qubus mode from a beam
splitter into a second mode that represents the environment. In
this case, the controlled displacements $D(\beta\sigma_z)$ can be
described as acting upon both the qubus mode,
$D_1(\sqrt{1-\eta^2}\beta\sigma_z)$, and the loss mode,
$D_2(\eta\beta\sigma_z)$, where $\eta$ is the reflectivity
parameter. The first observation is that the controlled
displacements on the qubus mode are no longer exactly those
required, leading to a smaller phase shift and an error in the
gate. It is also possible that the qubus mode will not disentangle
exactly from the qubits, if the phase space loops the qubus
traverses do not quite close. As long as the degree of loss is
known, these two effects can be eliminated by increasing the
amplitude $\beta'$ of the controlled displacement such that
$\beta=\sqrt{1-\eta^2} \beta'$. The most important effect to
consider is therefore the controlled displacements acting on the
loss mode, which cause a dephasing effect on the two-qubit state.
This effect scales as $\eta^2\alpha^2 \sin^2\theta$ and thus for
$\eta$ small ($\eta \ll 1$) this dephasing effect is minimal
[recall $\beta\simeq O(1)$].

In conclusion, we have demonstrated how to implement universal
two-qubit gates using fundamental atom-light interactions in
quantum optics, through qubus-mediated qubit-qubit communication
and without direct interaction between the qubits. In this hybrid
scheme, the only required interactions lead to controlled
rotations of a continuous-variable qubus mode, conditioned on the
state of the qubits. Our scheme is universal in the sense that any
regime is allowed for the controlled rotations, including
interactions in the limit of weak or strong nonlinearities. The
resulting phase gate is deterministic and measurement-free, and
thus represents a promising approach to implementing quantum
logic.

\noindent {\em Acknowledgments}: We thank R. van Meter and P. L.
Knight for valuable discussions. This work was supported in part
by MEXT in Japan, the EU project QAP and the ARC in Australia. SLB
currently holds a Royal Society Wolfson Research Merit Award.

% \vskip -0.5cm


\begin{thebibliography}{99}
\bibitem{footnoteatom} The atomic system could be a two-level atom or an effective
two-level system with an auxillary level ($\Lambda$-system).
\bibitem{footnote2} The qubus mode can be prepared in an arbitrary state, which
could even be mixed. However, the qubus should not evolve independently on timescales
comparable with that of the gate, for example, by decoherence.
\bibitem{signchange} The sign change here can be realised by a simple
bit flip on the matter qubit.
\bibitem{purecontrolleddispalcement}
The sequence of operations $U(\epsilon,\chi \sigma_z)$,
$U(\epsilon,-\chi \sigma_z)$, $U(-\epsilon,-\chi \sigma_z)$,
$U(-\epsilon,\chi \sigma_z)$ would achieve the controlled
displacement alone.
\end{thebibliography}
\end{document}